\RequirePackage{fix-cm}

\documentclass[smallextended]{svjour3}      
\smartqed  
\usepackage{graphicx}
\usepackage{here}
\usepackage{eurosym}
\usepackage{graphicx}
\usepackage{amsmath}
\usepackage{amssymb}
\usepackage{latexsym}
\usepackage{color}
\usepackage{epstopdf}
\usepackage[latin1]{inputenc}
\usepackage[french,english]{babel}
\usepackage[T1]{fontenc}

 \usepackage{mathptmx}      
\usepackage{cite}

\usepackage[colorlinks,citecolor=blue,urlcolor=blue,bookmarks=false,hypertexnames=true,allcolors=blue]{hyperref}

\usepackage{setspace} 
\usepackage{float}

\begin{document}

\title{Quantum key distribution using single photon added-subtracted squeezed coherent state}

\author{Y. Oulouda         \and
        M. El Falaki \and M. Daoud
}

\institute{Y. Oulouda \at
              Faculty of Sciences, Mohammed V University of Rabat, Morocco \\
              Tel.: +212-676-881432\\
              \email{youssef\_oulouda@um5.ac.ma}          
           \and
           M. El Falaki \at
              Faculty of Sciences Choua\"{\i}b Doukkali
              El Jadida, Morocco
              \and M. Daoud \at Faculty of Sciences Ibn Tofail,
              K\'enitra, Morocco
}

\date{Received: date / Accepted: date}

\maketitle

\begin{abstract}
	\par In this paper we investigate the security of continuous variable BB84 quantum key distribution protocol using single photon added then subtracted squeezed coherent state (SPASSCS). It's found that the SPASSCS is a non-Gaussian and non-classical state. Its non-Gaussianity and non-classicality are exhibited via the Wigner function. It's shown that the proposed state is generally robust against the eavesdropping strategies, such as intercept-resend attack and superior channel attack. Further, a comparative study has proved the strong efficiency of the proposed state over coherent state, squeezed coherent state and photon added then subtracted coherent state. Our analysis employs bit error rate, mutual information, and secure key gain.

	\keywords{Quantum key distribution, Secure key gain, Bit error rate}
\end{abstract}


\section{Introduction}
Quantum key distribution, is an encryption method that uses the laws of quantum mechanics to securely share a private key \cite{1,2}. This is due to a no-cloning theorem that prevents any attempts to duplicate the secret key, and measurement postulate that reveals any act of measurement from a cryptosystem to the legitimate users.\\

Various applications and researches have been made in hope to discover a solution to secure the key against eavesdropping; however, some of them suffer from inefficiency in long-distance transmission \cite{8}, sensitivity to quantum noise channel \cite{3}, and non-availability of the necessary equipment (repeaters/ unique photon sources).

Continuous variable quantum key distribution \cite{3} has attracted a lot of researches and rapid progress during this last few years \cite{4,5,25} and many applications have been achieved confirming their performance and efficient in many aspects, such as the availability of the equipment required to realize it experimentally, and also the possibility of developing a quantum state to become strong on average against eavesdropping and not dissipative in noisy quantum channel \cite{6}. Furthermore, continuous variable QKD using Gaussian states has played an important role in long distance transmission, with $202.81 km$ over an optic fibers \cite{8,9}, and $23 km$ in air \cite{10,11}. However, a Gaussian state has encountered some issues, such us the missing of large parts of the key, due to their sensitivity to noise channel. To overcome this obstacle a non Gaussian and nonclassical state has been proposed \cite{12}. Those properties were the subject of large works \cite{13,14,15}, which in turn moved into application field in quantum information science \cite{16,17}.\\

The aim of this work is to investigate the performance of continuous variable QKD using \textbf{S}ingle \textbf{P}hoton \textbf{A}dded then \textbf{S}ubtracted \textbf{S}queezed \textbf{C}oherent \textbf{S}tate SPASSCS as an efficient state to implement in QKD protocol. The idea engenders from Horak's work \cite{6} who successively proved the effect of squeezing operation in QKD, and Borelli's work who mentioned the effect of adding then subtracting photons to coherent state in QKD \cite{12}. We try to combine the two operations to create a robust state against eavesdropping attacks and quantum channel noise.

The paper is thus organized as follows. In section II, we investigate the non-Gaussianity and non-classicality of SPASSCS, by measuring the associate Wigner function. In section III, we describe the QKD scheme based on continuous variables BB84 protocol \cite{19} using SPASSCS in ideal conditions. We measure some relevant basic quantities such as probability distribution, post-selection efficiency, and bit error rate. In section IV, two classes of eavesdropping strategies are investigated. The first one is the intercept-resend attack, where an eavesdropper (Eve) performs a simultaneous measurements attack on both emerging signals from 50:50 beam splitter, then resends the signal to Bob according to her measurements. Second one is the superior channel attack, where Eve performs her measurement only on the reflected part of the signal sent by Alice, while the transmitted part is left to Bob. To study those attacks, some significant quantities are measured, joint probability, mutual information, bit error rate, and other functions. Finally, the results are summarized in section V.

\section{Single photon added-subtracted squeezed coherent state \label{sec2}}

\begin{figure}[H]
	\centering
	\includegraphics[scale=0.3]{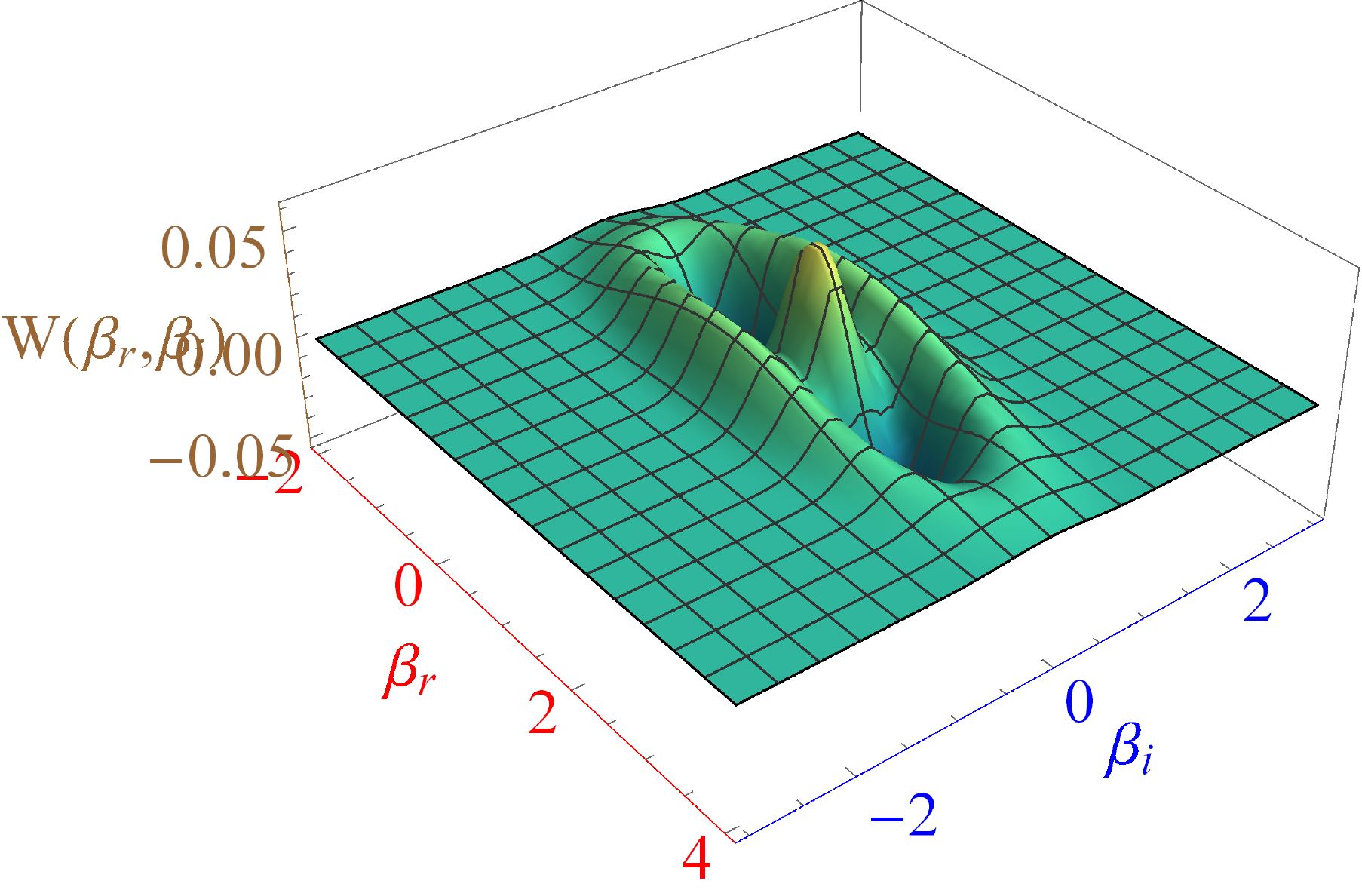}\hfill
	\includegraphics[scale=0.3]{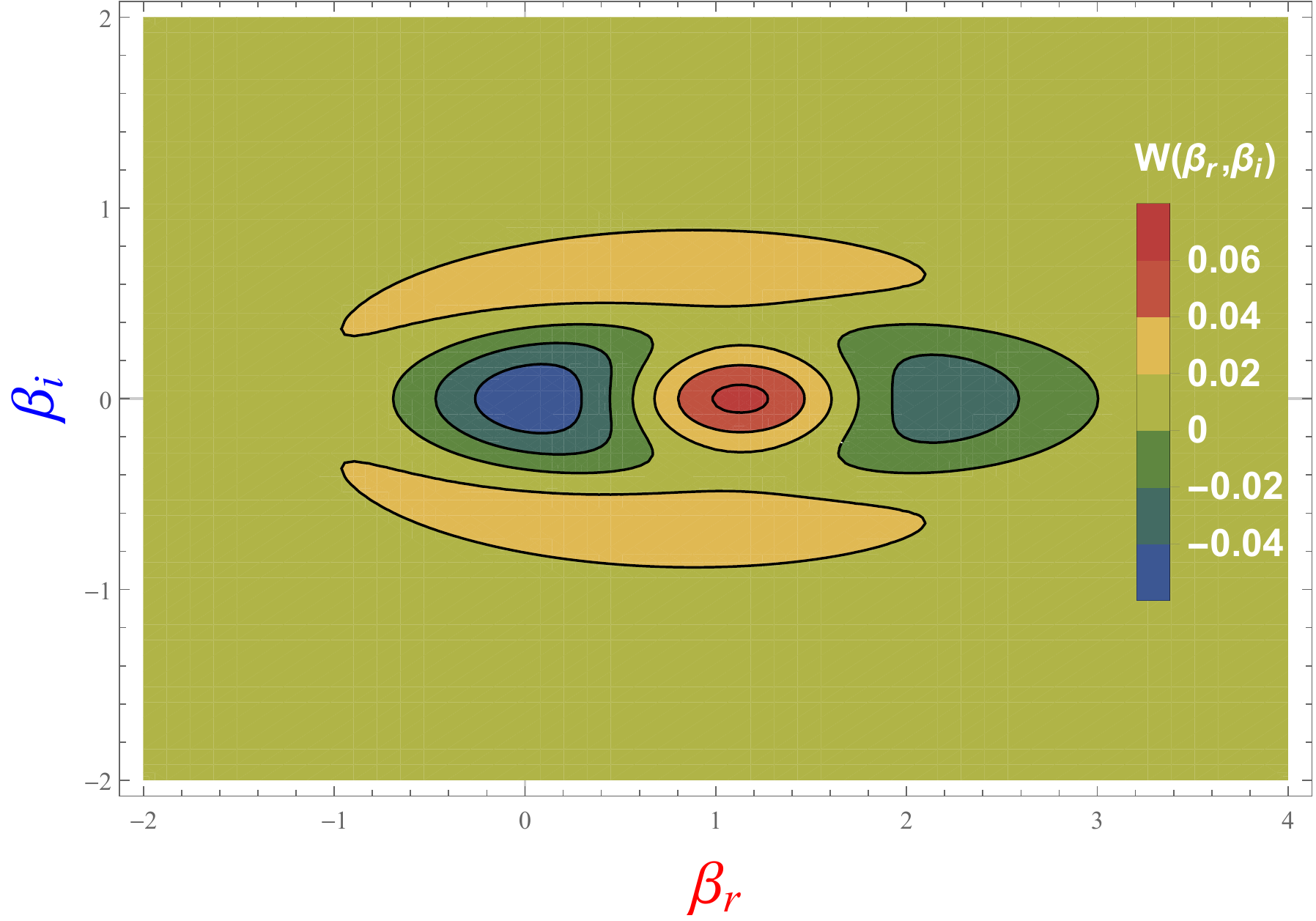}
	\caption{Wigner function of single photon added then subtracted squeezed coherent state $\alpha=1$ and $r=0.5$. In the left hand the three dimension representation of Wigner function, in the right hand, the contour plot of Wigner function.}
	\label{fig1}
\end{figure}

The single photon added-subtracted squeezed coherent state is obtained by adding first then subtracting one photon to squeezed coherent state
\begin{equation}
\displaystyle \left| \psi \right\rangle  = {{\mathcal{N}}^{ - \frac{1}{2}}}{a}{a^{ \dagger}}\left| SCS \right\rangle
\end{equation}
where ${{\mathcal{N}}^{ - \frac{1}{2}}}$ is normalization constant, and
\begin{equation}
\displaystyle \left|SCS \right\rangle  = {exp[\frac{r}{2}({a^{\dagger2}-a^2})]}{exp({\alpha a^{\dagger}-\alpha a})}\left| 0 \right\rangle .
\end{equation}

is the squeezed coherent state, $\alpha$ and $r$ are the coherent and the squeezed states parameters respectively, $\left| 0 \right\rangle$ is the
vacuum state. ${a}$, ${a^{ \dagger}}$ are Bose creation and annihilation operators, respectively, satisfying $[a,a^\dagger] = 1$.\\
One can prove the properties of this state by measuring their Wigner function which is a well-known indicator of a non-classicality \cite{21}.
\begin{equation}\label{9}
\displaystyle W\left( {\beta,\alpha,r } \right) = {e^{2{{\left| \beta  \right|}^2}}}\int_{ - \infty }^{ + \infty } {\frac{{{d^2}z}}{{{\pi ^2}}}\left\langle { - z} \right|\rho\left( {\alpha,r } \right) \left| z \right\rangle {e^{ - 2({\beta ^*}z - \beta {z^*})}}}.
\end{equation}
where  $\rho\left( {\alpha,r } \right)$ is the density operator of the state $\rho = \left| \psi \right\rangle \left\langle \psi \right|$, and $\beta={{\beta}_r}+i{{\beta}_i}$. The SPASSCS satisfies the nonclassical and non Gaussian properties. The non Gaussian property manifests in the form of three dimension representation of Wigner function in figure (\ref{fig1}), and the nonclassical property, manifests on the negative values of the Wigner function which can be seen clearly in contour plot representation in figure (\ref{fig1}). For additional information about those properties, one can consult \cite{13}.

\section{Quantum key distribution in continuous variables \label{sec3}}
Over the last few years, more attention has attracted the continuous variable quantum key distribution. This is due to rapid progress in many aspects, by avoiding most defects encountered the discrete variables quantum key distribution protocols.
The current section will explain in detail the continuous variables BB84 protocol using SPASSCS via homodyne detection and post-selection in absence of eavesdroppers. We tend to introduce some important quantities, density operator, post-selection efficiency and bit error rate.   
\subsection{Protocol description using SPASSCS}
Alice prepares the single-photon added then subtracted squeezed coherent state with field amplitude $\alpha$ and squeezing parameter $r$. We assume that both $\alpha$ and $r$ are real numbers. Where $r>0$ denotes an amplitude squeezed state, $r\le 0$ denotes a phase squeezed state, while $r=0$ denotes a single photon added then subtracted coherent state \cite{12,20}. Alice creates her encoding states by shifting the pulse phase $\psi(r,\alpha e^{i\varphi})$ as follows
\begin{eqnarray}
\centering	
B_1=\left\{
\begin{array}{r c l l}
\varphi=&0& \hspace{0.5cm}{\left| \psi_{+}  \right\rangle}:=&\psi(r,\alpha)\\
\varphi=&\pi&\hspace{0.5cm}{\left| \psi_{-}  \right\rangle}:=&\psi(r,\alpha e^{i\pi}) \nonumber
\end{array}
\right.,
\end{eqnarray}
\begin{eqnarray}		
\displaystyle B_2=\left\{
\begin{array}{r c l l}
\varphi=&\frac{\pi}{2}&\hspace{0.5cm}{\left| \psi_{+i}  \right\rangle}:=&\psi(r,\alpha e^{i\frac{\pi}{2}})\\
\varphi=&\frac{3\pi}{2}&\hspace{0.5cm}{\left| \psi_{-i}  \right\rangle}:=&\psi(r,\alpha e^{i\frac{3\pi}{2}}) \nonumber
\end{array}
\right.
\end{eqnarray}
After the construction of the coding bases $B_1$ and $B_2$. Alice encodes her bit sequence "0" and "1" within the four states randomly. Then she sends the encoding states through a quantum channel to Bob.\\
{\begin{minipage}[b]{0.5\linewidth}
		\centering	
		\[
		"0"=\left\{
		\begin{array}{r c l}
		&{\left| \psi_{-}  \right\rangle}&\\
		&{\left| \psi_{-i}  \right\rangle}&\nonumber
		\end{array}
		\right.;
		\]
	\end{minipage}\hfill
	\begin{minipage}[b]{0.5\linewidth}
		\centering
		\[
		"1"=\left\{
		\begin{array}{r c l}
		&{\left| \psi_{+}  \right\rangle}&\\
		&{\left| \psi_{+i}  \right\rangle}&\nonumber
		\end{array}
		\right.
		\]
\end{minipage}}\\
Bob in his part prepares to receive the encoding pulses sent from Alice with his two quadratures measurements ${\hat z}_1$ and ${\hat z}_2$. We define the quadratures measurements by the relation $\hat a={\hat z}_{1} + i {\hat z}_{2}$ and ${\hat a}^{ \dagger}={\hat z}_{1} - i {\hat z}_{2}$ thus $[{\hat z}_1 , {\hat z}_2]=\frac{i}{2}$. For each signal received he chooses randomly one of the quadratures measurements ${\hat z}_1$ or ${\hat z}_2$ with homodyne detection \cite{22} to make his measurement. After the transmission of a large number of pulses, Alice and Bob communicate through a classical channel for reconciliation. In this step, they divide the resulting data into correct-basis data and wrong-basis data according to Alice announcement. The pulse is in correct-basis data if Bob has chosen his quadrature measurement ${\hat z}_1$ when Alice sent ${\left| \psi_{\pm} \right\rangle}$ or if he has chosen ${\hat z}_2$ when Alice sent ${\left| \psi_{\pm i} \right\rangle}$. And the other projections, the pulse called wrong-basis data. For the correct-basis pulses, Bob sets the threshold value $z_c$ of his post-selection and selects his bits as follow
\begin{eqnarray}\label{zc}
\left\{
\begin{array}{r c l l}
 z_{1,2} < &-&z_c & \longrightarrow "0"\\
 z_{1,2} > & &z_c &\longrightarrow "1"\\
otherwise  & & &\longrightarrow neglect
\end{array}
\right.,
\end{eqnarray}

The gap $\left] { - {z_c},{z_c}} \right[$ represents the uncertainty interval, where the bit selected is unknown or non confirmed fact exists.The post-selection plays a crucial role in continuous variables QKD protocols \cite{6,19,24}.

\subsection{Basic quantities and bit error rate}
The aim of this section is to give a mathematical interpretation of the protocol using SPASSCS. Alice sends one of the four states randomly with equal probability, thus the density operator is described by
\begin{equation}\label{123}
\displaystyle \rho= \frac{1}{4}\left( {\left| {{\psi _ + }} \right\rangle \left\langle {{\psi _ + }} \right| + \left| {{\psi _ - }} \right\rangle \left\langle {{\psi _ - }} \right|} +
{\left| {{\psi _{ + i}}} \right\rangle \left\langle {{\psi _{ + i}}} \right| + \left| {{\psi _{ - i}}} \right\rangle \left\langle {{\psi _{ - i}}} \right|} \right).
\end{equation}

\begin{figure}[H]
	{\begin{minipage}[b]{0.48\linewidth}
			\centering
			\includegraphics[scale=0.3]{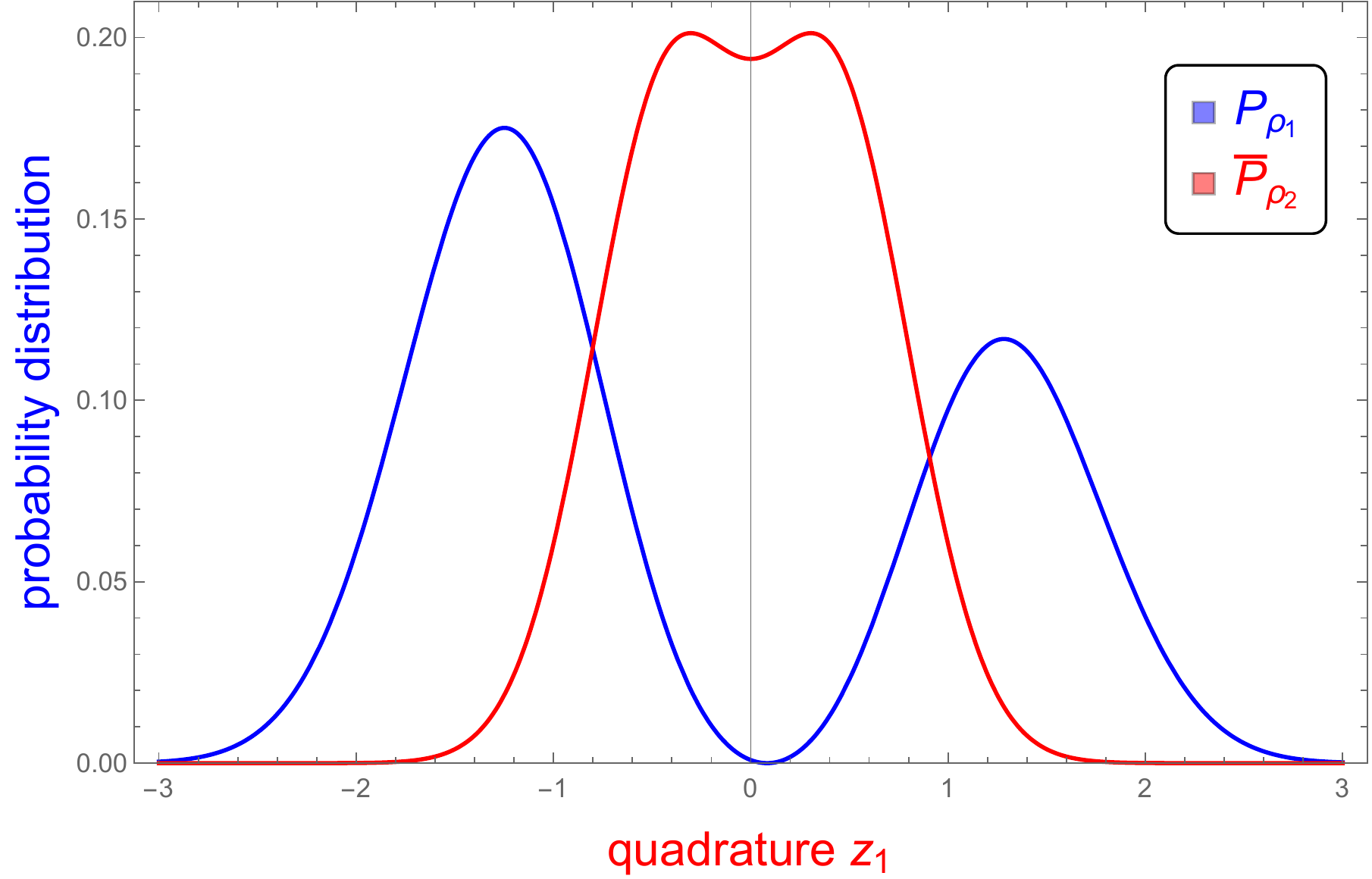}\\a

			\vfill     
		\end{minipage}\hfill
		\begin{minipage}[b]{0.48\linewidth}
			\centering
			\includegraphics[scale=0.3]{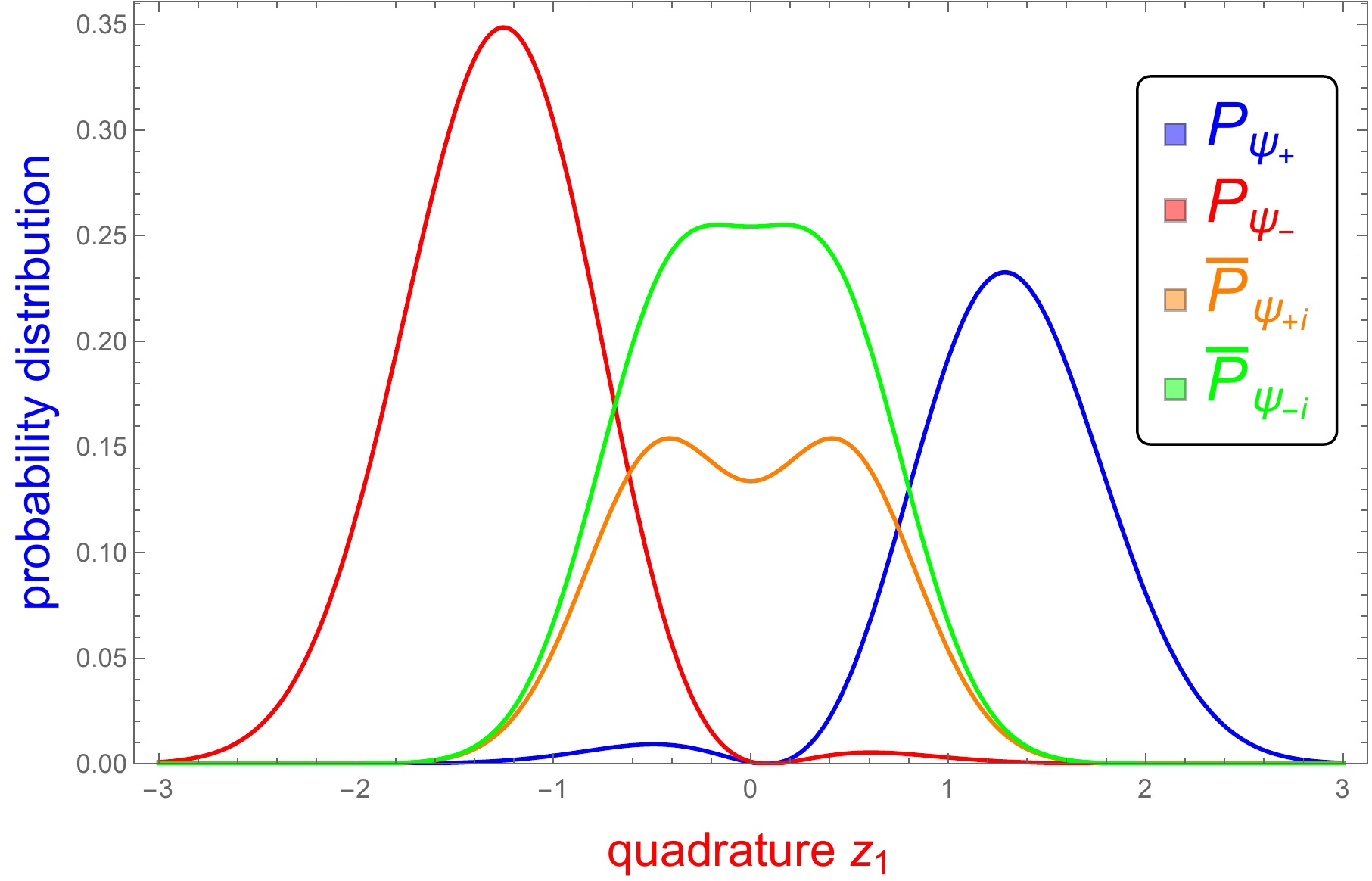}\\b
			
					\vfill     
	\end{minipage}

\caption{\textbf{a}: Probabilities distributions of correct-basis data (blue-line) and wrong-basis data (red-line). \textbf{b}: Probabilities distributions of states projections in correct-basis data: (red-line stands for $\left| {{\psi _ - }} \right\rangle$ and blue-line for $\left| {{\psi _ + }} \right\rangle$) and states projections in wrong-basis data: (green-line stands for $\left| {{\psi _ {-i} }} \right\rangle$ and orange-line for $\left| {{\psi _ {+i} }} \right\rangle$), where $\alpha=1$ and $r=0.2$ in absence of eavesdroppers.}\label{cbwbd}}
\end{figure}

\begin{figure}[H]
	\centering
	\includegraphics[scale=0.35]{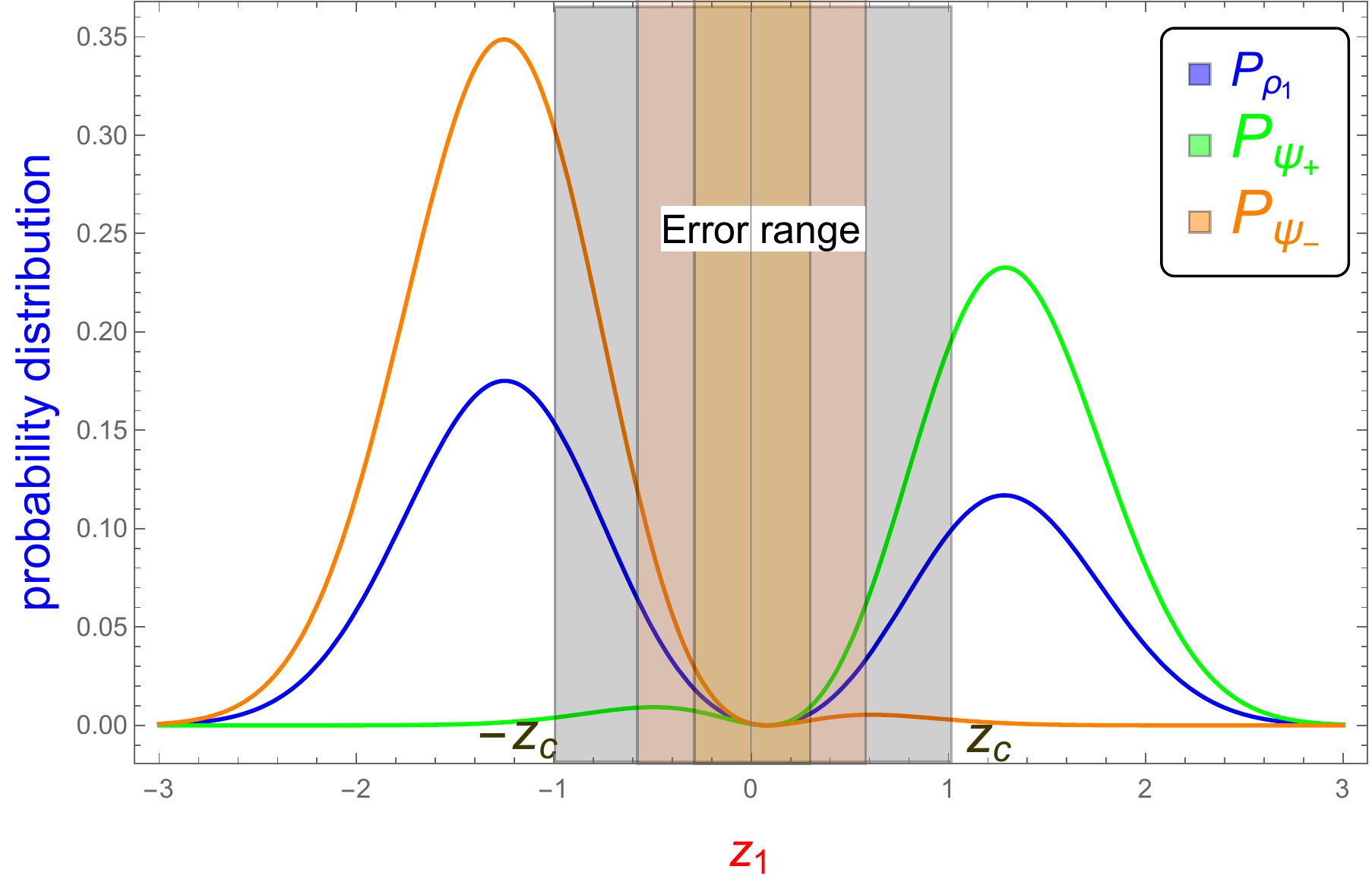}
	\caption{Possibles error ranges in the absence of eavesdroppers, where $\alpha=1$ and $r=0.2$.}
	\label{errange}
\end{figure}

Note that, in order to have a secure communication, the coding states must not be orthogonal to each other, so Eve -the eavesdropper- would be unable to distinguish between the states, and by coincidence revealing the secret key bit. In our case, the states are not orthogonal between them, so complete differentiation is impossible. In the reconciliation step, Alice announces the quadrature measurements $z_1$ and $z_2$ (the bases) on which she encoded her bit sequence, so the density operator defined in (\ref{123}) is simplified into

\begin{eqnarray}\label{rr}
\left\{\begin{array}{l}
\displaystyle {\rho _1} = \frac{1}{2}\left( {\left| {{\psi _ + }} \right\rangle \left\langle {{\psi _ + }} \right| + \left| {{\psi _ - }} \right\rangle \left\langle {{\psi _ - }} \right|} \right)\\
\displaystyle {\rho _2} = \frac{1}{2}\left( {\left| {{\psi _{ + i}}} \right\rangle \left\langle {{\psi _{ + i}}} \right| + \left| {{\psi _{ - i}}} \right\rangle \left\langle {{\psi _{ - i}}} \right|} \right)
\end{array}\right.
\end{eqnarray}
The bases announcement discloses the correct-basis data as well as the wrong-basis data. Bob's final challenge is to find out the right bit, which is the post-selection's assignment (\ref{zc}).\\
The mathematical word of correct-basis and wrong-basis results are represented in probability distribution measurement. For correct-basis distribution
\begin{eqnarray}\label{key}
\begin{array}{l}
\displaystyle {P_{{\rho _1}}}\left( {{z_1},r,\alpha } \right) = \frac{1}{2}\left( {{P_{{\psi _ + }}}\left( {{z_1},r,\alpha } \right) + {P_{{\psi _ - }}}\left( {{z_1},r,\alpha } \right)} \right)\\
\displaystyle {P_{{\rho _2}}}\left( {{z_2},r,\alpha } \right) = \frac{1}{2}\left( {{P_{{\psi _ {+i} }}}\left( {{z_2},r,\alpha } \right) + {P_{{\psi _ {-i} }}}\left( {{z_2},r,\alpha } \right)} \right)
\end{array}
\end{eqnarray}
and for wrong-basis distribution 
\begin{eqnarray}\label{key}
\begin{array}{l}
\displaystyle {{\bar P}_{{\rho _1}}}\left( {{z_2},r,\alpha } \right) = \frac{1}{2}\left( {{{\bar P}_{{\psi _{ + }}}}\left( {{z_2},r,\alpha } \right) + {{\bar P}_{{\psi _{ - }}}}\left( {{z_2},r,\alpha } \right)} \right)\\
\displaystyle {{\bar P}_{{\rho _2}}}\left( {{z_1},r,\alpha } \right) = \frac{1}{2}\left( {{{\bar P}_{{\psi _{ + i}}}}\left( {{z_1},r,\alpha } \right) + {{\bar P}_{{\psi _{ - i}}}}\left( {{z_1},r,\alpha } \right)} \right)
\end{array}
\end{eqnarray}
Figure (\ref{cbwbd}-\textbf{a}) depicts those two distributions as a function of $z$. Where
\begin{eqnarray}\label{key}
\begin{array}{l}
\displaystyle {P_{{\psi _ \pm }}}\left( {{z_1},r,\alpha } \right) ={\left| {\left\langle {{{z_1}}}
		\mathrel{\left | {\vphantom {{{z_1}} {{\psi _ \pm }}}}
			\right. \kern-\nulldelimiterspace}
		{{{\psi _ \pm }}} \right\rangle } \right|^2} = \mathop \int_{ - \infty }^{ + \infty} W\left( {{z_1},{z_2},r, \pm \alpha } \right){\rm{d}}{z_2}\\
\displaystyle {P_{{\psi _{ \pm i}}}}\left( {{z_2},r,\alpha } \right) ={\left| {\left\langle {{{z_2}}}
		\mathrel{\left | {\vphantom {{{z_2}} {{\psi _{ \pm i}}}}}
			\right. \kern-\nulldelimiterspace}
		{{{\psi _{ \pm i}}}} \right\rangle } \right|^2}  = \mathop \int_{ - \infty }^{ + \infty} W\left( {{z_1},{z_2},r, \pm i\alpha } \right){\rm{d}}{z_1}
\end{array}
\end{eqnarray}
are the probability distributions of correct-basis states projections and
\begin{eqnarray}\label{key}
\begin{array}{l}
\displaystyle {{\bar P}_{{\psi _{ \pm i}}}}\left( {{z_1},r,\alpha } \right) ={\left| {\left\langle {{{z_1}}}
		\mathrel{\left | {\vphantom {{{z_1}} {{\psi _{ \pm i}}}}}
			\right. \kern-\nulldelimiterspace}
		{{{\psi _{ \pm i}}}} \right\rangle } \right|^2} = \mathop \int_{ - \infty }^{ + \infty} W\left( {{z_1},{z_2},r, \pm i\alpha } \right){\rm{d}}{z_2}\\
\displaystyle {{\bar P}_{{\psi _ \pm }}}\left( {{z_2},r,\alpha } \right) ={\left| {\left\langle {{{z_2}}}
		\mathrel{\left | {\vphantom {{{z_2}} {{\psi _ \pm }}}}
			\right. \kern-\nulldelimiterspace}
		{{{\psi _ \pm }}} \right\rangle } \right|^2} = \mathop \int_{ - \infty }^{ + \infty} W\left( {{z_1},{z_2},r, \pm \alpha } \right){\rm{d}}{z_1}
\end{array}
\end{eqnarray}
are the probability distributions of wrong-basis states projections. The representation of those states distributions are depicted in Figure (\ref{cbwbd}-\textbf{b}).
In the reconciliation step when the legitimate users communicate through a public channel to compare their data; Alice announcement of the bases results to the correct-basis and wrong-basis distributions figure (\ref{cbwbd}-\textbf{a}), and Alice announcement of the states results to the state projection distributions of the correct-basis and wrong-basis data figure (\ref{cbwbd}-b). However, Alice's revelation of the states compromised the entire protocol by exposing the secret key bit to prospective eavesdroppers.\\
 To prevent this from happening, she commits to simply announcing the bases, while Bob uses his post-selection to extract the correct states. The error range occurs exclusively between states projection distributions ${\left|{\left\langle{{{z_1}}}\mathrel{\left | {\vphantom {{{z_1}} {{\psi _ + }}}}\right. \kern-\nulldelimiterspace}{{{\psi _ + }}} \right\rangle } \right|^2}$ and ${\left|{\left\langle{{{z_1}}}\mathrel{\left | {\vphantom {{{z_1}} {{\psi _ - }}}}\right. \kern-\nulldelimiterspace}{{{\psi _ - }}} \right\rangle } \right|^2}$ of a correct-basis data, which take place around $z_1 =0$ figure (\ref{errange}). For this issue Bob sets his threshold value $z_c$ to ignore this inconclusive range and having non doubt in the final results.
To avoid this error range, we must have an efficient post-selection capable to set a certain bit value "0" or "1" from correct-basis data. For this issue, we define the post-selection efficiency $\Pi \left( {{z_c},r,n} \right)$ as the probability of post-selection to reveal the carried bit encoded in the pulse characterized by a quadrature measurement $z_{1}$ or $z_{2}$.
\begin{equation}\label{key}
\displaystyle \Pi \left( {{z_c},r,n} \right) = \int_{ - \infty }^{ - {z_c}} {{P_{{\rho _1}}}\left( {{z_1},r,\sqrt n } \right)d{z_1}}   + \int_{{z_c}}^{ + \infty } {{P_{{\rho _1}}}\left( {{z_1},r,\sqrt n } \right)d{z_1}} 
\end{equation}

Where $n=\alpha^2$ is the pulse intensity (the mean photon number per pulse). We define the bit error rate BER for a given $z_c$ as the probability of Bob occurs an error, i.e. the probability that the post-selection result to $z_1<-z_c$ when Alice sends $\left| {{\psi _ + }} \right\rangle $
\begin{equation}\label{key}
\displaystyle \Theta\left( {{z_c},r,n} \right) = \frac{1}{{\Pi\left( {{z_c},r,n} \right)}}\int_{ - \infty }^{ - {z_c}} {{P_{{\psi _ + }}}\left( {{z_1},r,\alpha } \right)d{z_1}} 
\end{equation}
\begin{figure}[H]
	\centering
	\includegraphics[scale=0.35]{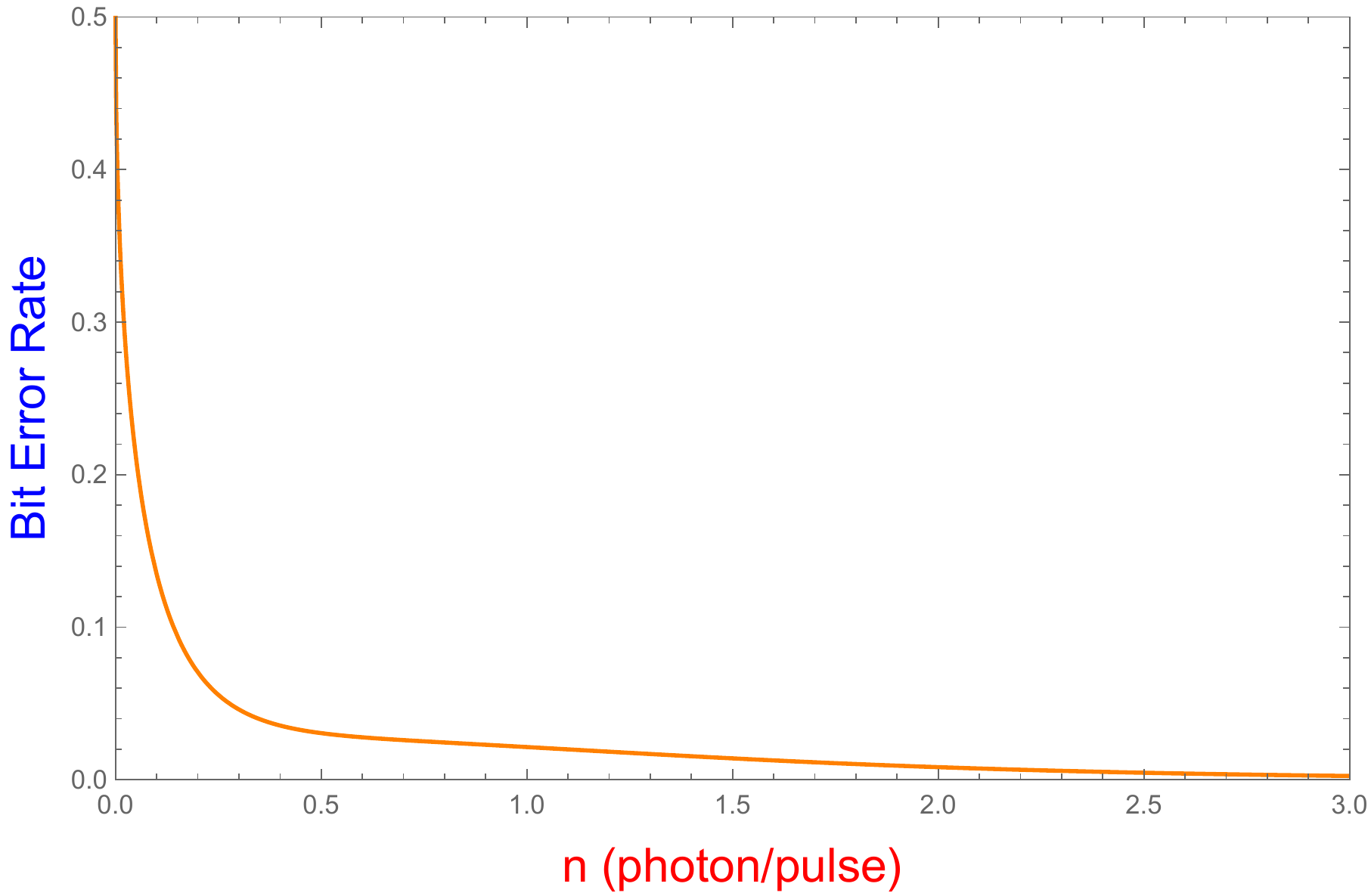}
	\caption{Bit error rate of Bob measures bit "0" when Alice sent $\left| {{\psi _ + }} \right\rangle $ for $z_c=0$, $\alpha=\sqrt n$ and $r=0.2$ in the absence of eavesdroppers.}
	\label{berwe}
\end{figure}
To obtain a lower bit error rate we decrease the post-selection efficiency by increasing the threshold value $z_c$, because $\Pi$ and $\Theta$ are decreasing functions.
The security of the entire system depends on choosing an appropriate $z_c$ and $\alpha$ values, which will be the subject of the next section.

\section{Eavesdropping strategies}
There is no ideal communication during the transmission of a private key. The signal is either absorbed by the environment or intercepted by an eavesdropper. In both cases, the transmission is simulated by a beam-splitter characterized by an amplitude transmission $T$ and reflection $R$, where ${T^2}+{R^2}=1$. Here we assume a 50:50 beam splitter, meaning $T^2=R^2=\frac{1}{2}$.
\subsection{Intercept resend attack}
In intercept-resend attack, Eve measures the whole signal sent by Alice then resends it to Bob according to her measurement results. We are interested to study the intercept resend attack based on simultaneous measurement of both quadratures components of beam-splitter \cite{6}.

 Eve measures the two emerging beams of the beam-splitter simultaneously with $z_1$ and $z_2$.
Then she makes her bit sequence results according to this comparison.
\begin{equation}\label{key}
\left\{ {\begin{array}{*{20}{c c}}
	{\left| {{\psi _ + }} \right\rangle\hspace{1.5cm}} &if& {\hspace{0.2cm}{z_1} \ge \left| {{z_2}} \right|}\\
	{\left| {{\psi _{ + i}}} \right\rangle\hspace{1.5cm}} &if&{ \hspace{0.2cm}{z_2} > \left| {{z_1}} \right|}\\
	{\left| {{\psi _ - }} \right\rangle\hspace{1.5cm}} &if& {\hspace{0.2cm} - {z_1} \ge \left| {{z_2}} \right|}\\
	{\left| {{\psi _{ - i}}} \right\rangle\hspace{1.5cm}} &if& {\hspace{0.2cm} - {z_2} > \left| {{z_1}} \right|}
	\end{array}} \right.
\end{equation}

Let's assume for simplicity, that Alice sent the state $\left| {{\psi _ + }} \right\rangle$. Eve intercepts Alice signal by measuring simultaneously the two emerging beams of the beam-splitter, one output arm with quadrature measurement $z_1$, and the other output arm using $z_2$, the probability distribution of the intercepted signal is given by 
\begin{equation}\label{key}
\mathcal{ E} \left( {{z_1},{z_2},r,\alpha } \right) = {P_{{\psi _ + }}}\left( {{z_1},r,\frac{\alpha}{\sqrt{2}} } \right){{\bar P}_{{\psi _ + }}}\left( {{z_2},r,\frac{\alpha}{\sqrt{2}} } \right)
\end{equation}
Eve then will try to resend the signal to Bob based on her measurement. She has three options in this stage: she either resends the signal in its original state $\left| {{\psi _ + }} \right\rangle$ with probability
\begin{equation}\label{key}
{e_0}\left( {r,\alpha } \right) = \int_{{z_1} > \left| {{z_2}} \right|} {\mathcal{ E} \left( {{z_1},{z_2},r,\alpha } \right)d{z_1}d{z_2}} 
\end{equation}
or she resends a $\pi$-phase shifted signal $\left| {{\psi _ - }} \right\rangle$ with probability
\begin{equation}\label{key}
{e_\pi }\left( {r,\alpha } \right) = \int_{{z_2} > \left| {{z_1}} \right|} {\mathcal{ E} \left( {{z_1},{z_2},r,\alpha } \right)d{z_1}d{z_2}} 
\end{equation}
or she resends a $\frac{\pi}{2}$-phase shifted signal in wrong-basis status  $\left| {{\psi _ {+i} }} \right\rangle$ or $\left| {{\psi _ {-i} }} \right\rangle$ with equal probability
\begin{equation}\label{key}
{e_{\frac{\pi }{2}}}\left( {r,\alpha } \right) = \int_{ - {z_1} > \left| {{z_2}} \right|} {\mathcal{ E} \left( {{z_1},{z_2},r,\alpha } \right)d{z_1}d{z_2}} 
\end{equation}
As a result, the original state $\left| {{\psi _ + }} \right\rangle$ will be transformed into
\begin{equation}\label{aat}
\begin{array}{l}
\left| {{\psi _ + }} \right\rangle \left\langle {{\psi _ + }} \right| \to 
{e_0}\left| {{\psi _ + }} \right\rangle \left\langle {{\psi _ + }} \right| + {e_\pi }\left| {{\psi _ - }} \right\rangle \left\langle {{\psi _ - }} \right|\\
+ {e_{\frac{\pi }{2}}}\left( {\left| {{\psi _{ + i}}} \right\rangle \left\langle {{\psi _{ + i}}} \right| + \left| {{\psi _{ - i}}} \right\rangle \left\langle {{\psi _{ - i}}} \right|} \right)
\end{array}
\end{equation}
And the densities operators $\rho {_1}$ and $\rho {_2}$ (\ref{rr}) after Alice announcement of the bases become
\begin{equation}\label{key}
\begin{array}{l}
\rho {'_1} = \left( {{e_0} + {e_\pi }} \right){\rho _1} + 2{e_{\frac{\pi }{2}}}{\rho _2}\\
\rho {'_2} = 2{e_{\frac{\pi }{2}}}{\rho _1} + \left( {{e_0} + {e_\pi }} \right){\rho _2}
\end{array}
\end{equation}
Due to Eve's attack, the probability distribution of correct-basis and wrong-basis have been disturbed, and they are given by ${P_{{\rho' _1}}}\left( {{z_1},r,\alpha } \right)$ and ${{\bar P}_{{\rho' _2}}}\left( {{z_1},r,\alpha } \right)$ respectively
\begin{figure}[H]
	\centering
	\includegraphics[scale=0.35]{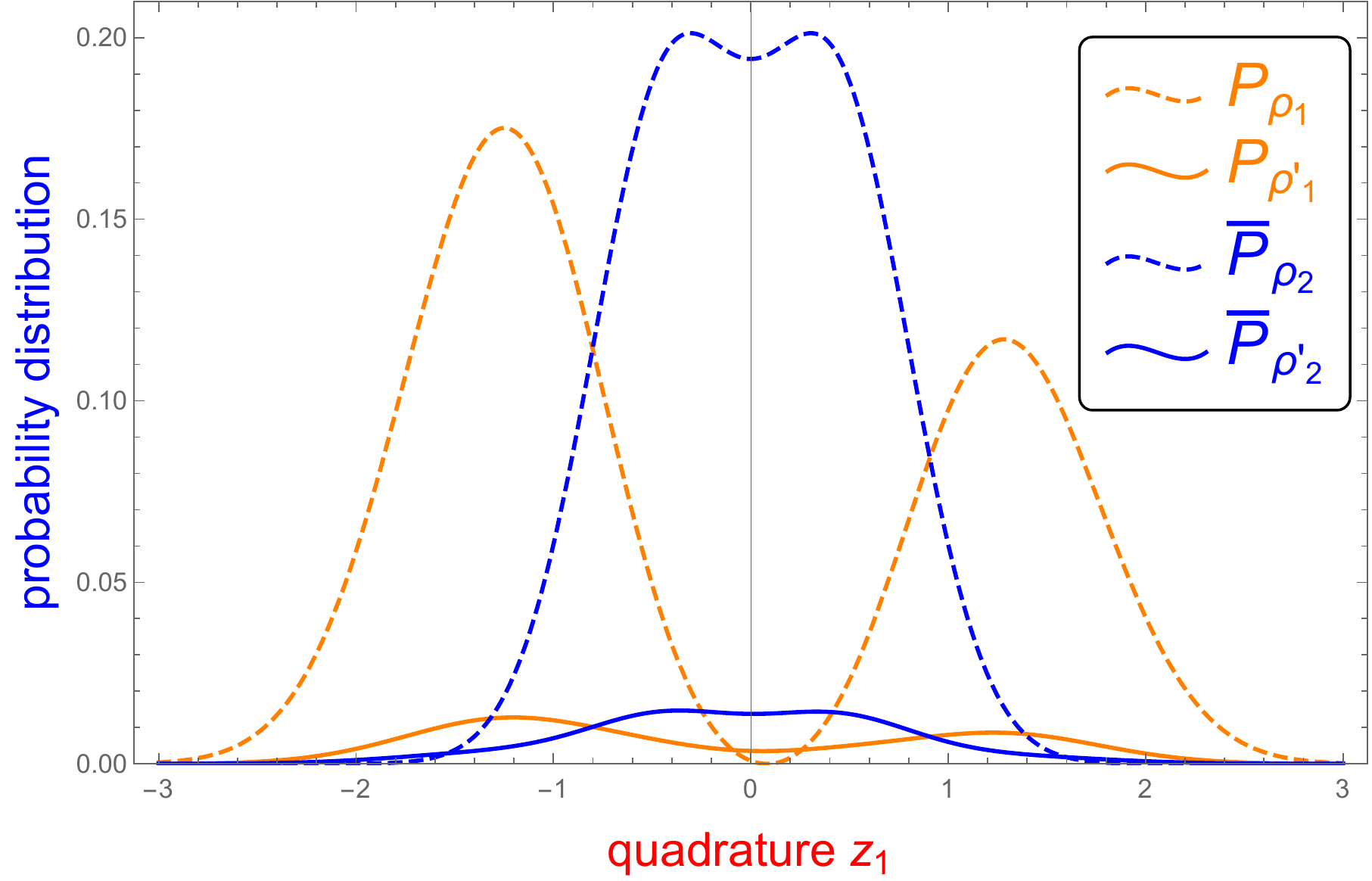}
	\caption{Probabilities distributions of correct-basis and wrong-basis data for $\alpha=1$ and $r=0.2$ in the absence and presence of eavesdroppers. The dashed lines stands for the absence of eavesdroppers, and solid lines for the presence of eavesdroppers. The orange lines represent the correct-basis data while the blues lines represent the wrong-basis data.}
	\label{dpd}
\end{figure}
The intervention of eavesdropper by mean of intercept-resend attack changes the form of the probability distribution of the bases figure (\ref{dpd}), which could give Bob an evidence of their existence, allowing him to either ignore the signal or perform error correction. For that reason, one should measure Bob bit error rate after attack

\begin{equation}\label{key}
\displaystyle \Theta{'}\left( {{z_c},r,n} \right) = \frac{1}{{\Pi{'}\left( {{z_c},r,n} \right)}}\int_{ - \infty }^{ - {z_c}} {{P_{{\psi{'_ + } }}}\left( {{z_1},r,\alpha } \right)d{z_1}} 
\end{equation}

where ${P_{{\psi{'_ + } }}}$ is the probability distribution of the state $\left| {{\psi _ + }} \right\rangle $ after attack shown in (\ref{aat}),  $\Pi{'} \left( {{z_c},r,n} \right)$ is the post-selection efficiency after intercept-resend attack given by
\begin{equation}\label{key}
\displaystyle \Pi{'} \left( {{z_c},r,n} \right) = \int_{ - \infty }^{ - {z_c}} {{P_{{\rho {'_1}}}}\left( {{z_1},r,\sqrt n } \right)d{z_1}}   + \int_{{z_c}}^{ + \infty } {{P_{{\rho {'_1}}}}\left( {{z_1},r,\sqrt n } \right)d{z_1}} 
\end{equation}
\begin{figure}[H]
	\centering
	\includegraphics[scale=0.35]{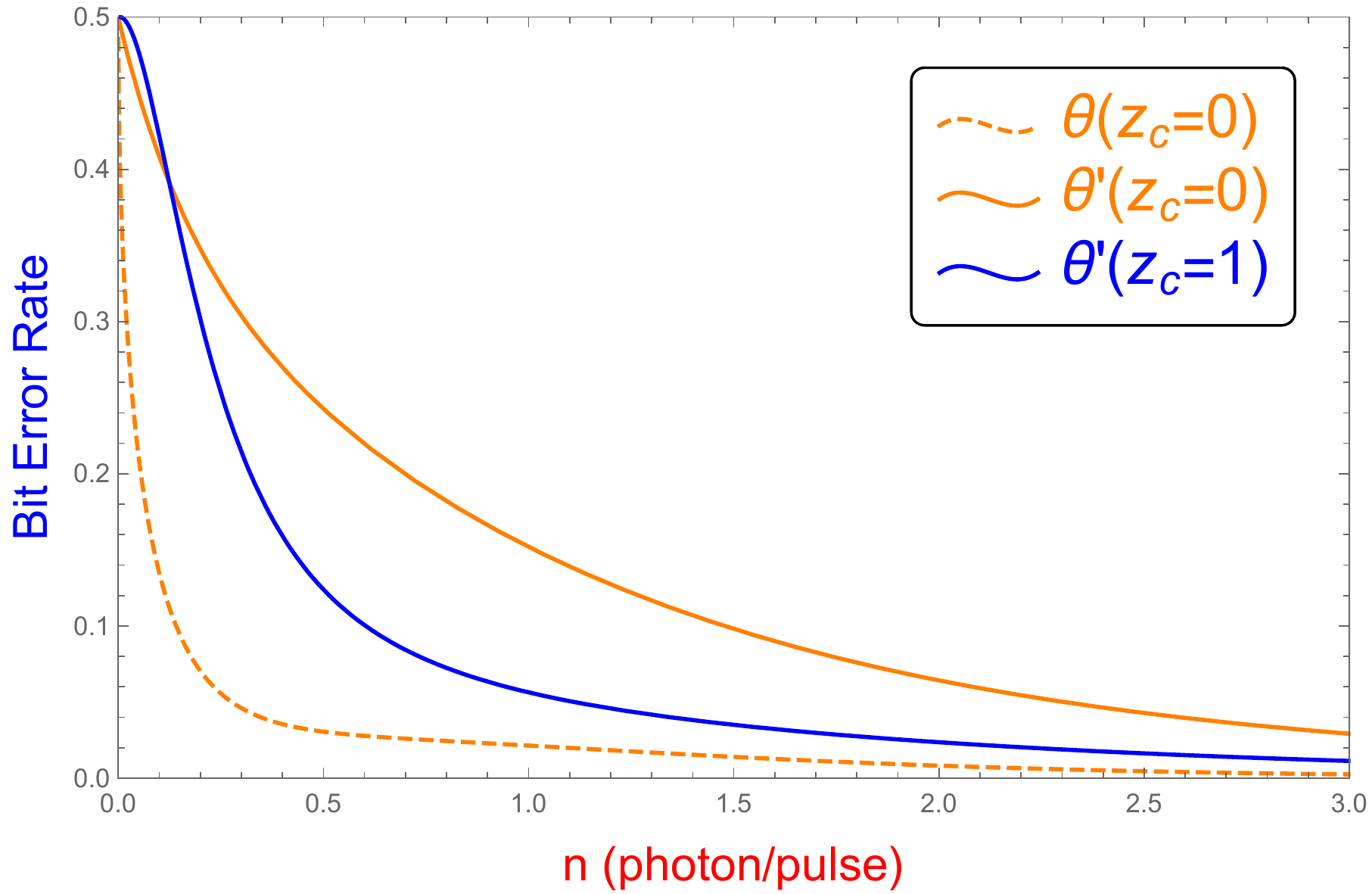}
	\caption{Bit error rate of Bob measures bit "0" when Alice sent $\left| {{\psi _ + }} \right\rangle $ for $\alpha=\sqrt n$ and $r=0.2$ after intercept-resend attack. The dashed line stands for the absence of eavesdropper and the solid lines for the presence of eavesdropper, while the orange lines (dashed and solid) represent the case where $z_c =0$, and blue line for $z_c =1$. }
	\label{beraa}
\end{figure}
Eve's intervention not only affects on the probability distribution of the bases figure (\ref{dpd}), but also affects on Bob's ability to occur errors in his measurement. The orange lines in figure (\ref{beraa}) clearly show how the bit error rate increases after Eve's intervention. 
But Bob can adjust the result in his favour, by raising the threshold value $z_c$, and decreasing by that the bit error rate, as the blue line shows in figure (\ref{beraa}).
Eve does not have this feature, because she does not have a post-selection to change the threshold value as she wants, her threshold value is always zero.
\subsection{Superior channel attack}
Quantum channel noise could be a beneficial support to eavesdroppers, Eve uses the channel imperfection on her favour. She is assumed to own a lossless channel and a quantum memory capable to store the encoding pulses until the reconciliation step. 

Eve stores the reflected part of the pulse emerging from 50:50 beam-splitter which would normally be lost in quantum channel, and lets the transmitted part intact via a lossless channel to Bob, then she waits for the bases announcement to perform her measurement on the stored pulses in correct-bases.\\
Unfortunately, this kind of attack cannot be detected by the legitimate users due to the indistinguishability from the ordinary quantum noise.\\
To measure the amount of information obtained by Bob and Eve during this attack, the joint probability distribution ${\mathcal{P}_ \pm }\left( {{z},{x}} \right)$ is proposed, in which Bob obtains the value $z$ of his quadrature measurement and Eve obtains the value $x$ of her quadrature measurement.

Let's assume for simplicity, that Alice sent the pulse state $\left| {{\psi _ + }} \right\rangle $, in which case Eve measures the reflected signal of the beam-splitter with $x_1$ and Bob measures the transmitted signal with $z_1$. Bob and Eve's joint probability distribution is given by

\begin{equation}\label{key}
\displaystyle {\mathcal{P}_{{\psi _{+}  }}}\left( {{z_1},{x_1},r,\alpha } \right) 
=\mathop  \int  W\left( {\frac{1}{\sqrt{2}}\left({{z_1}-{x_1}}\right),\frac{1}{\sqrt{2}}\left({{z_2}-{x_2}}\right),r, \alpha } \right){\rm{d}}{z_2}{\rm{d}}{x_2} 
\end{equation}
\begin{figure}[H]
	\centering
	\includegraphics[scale=0.2]{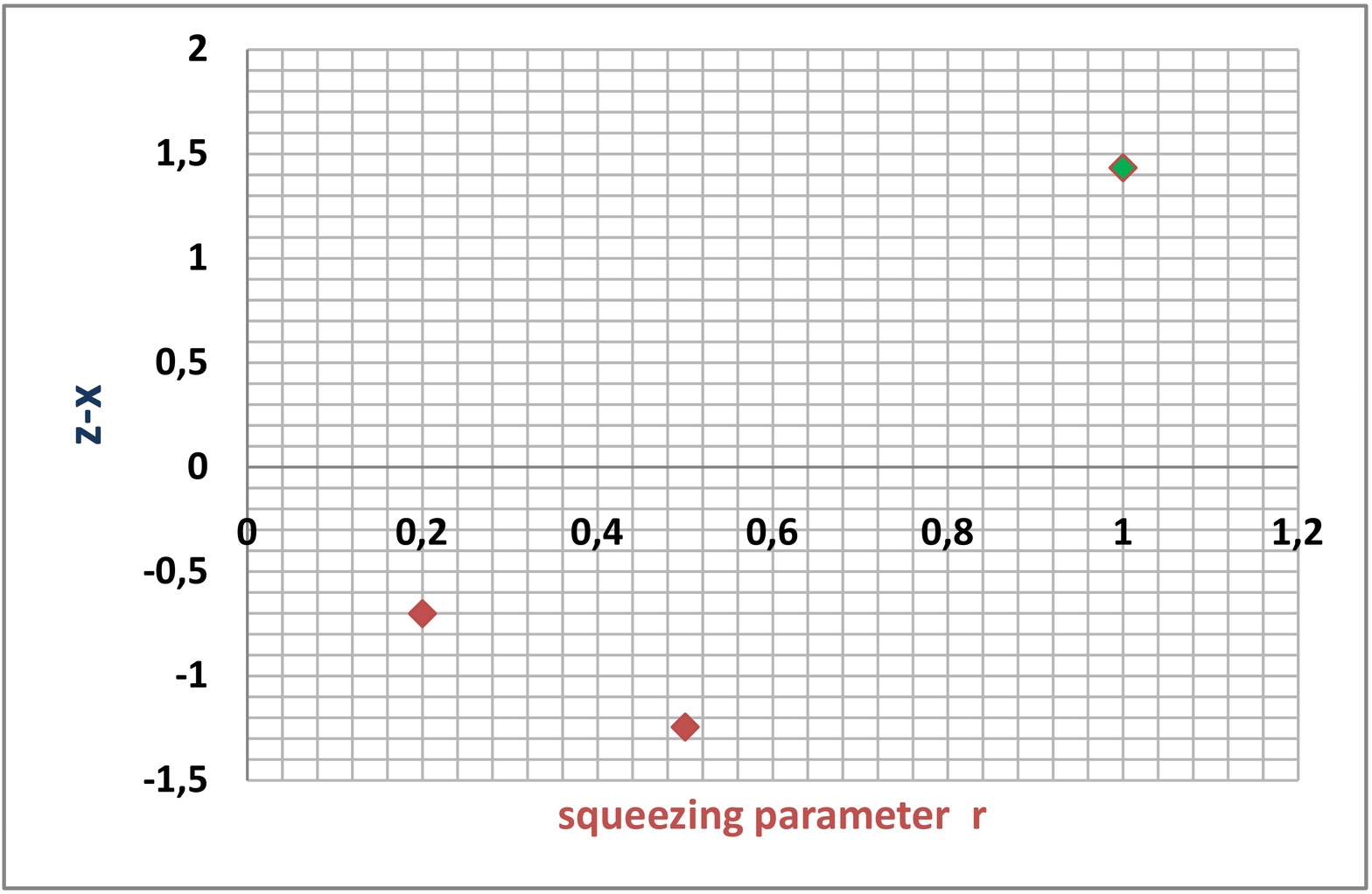}
	\caption{Comparative results between Bob and Eve measurement $z$ and $x$ coordinates respectively when the joint probability is maximum for different values of squeezing parameter $r$ with $\alpha=1$ and $T^{2}=R^{2}=\frac{1}{2}$.}
	\label{jpd}
\end{figure}

Since the joint probability of coherent state is factorizable, Bob's and Eve's measurements are independent, which make difficult to control Eve's measurement. However, the joint probability of squeezed state \cite{6} and PASCS \cite{12}, as well as the current state, are not factorizable, implying that Bob's and Eve's measurements are correlated, allowing Bob to influence on Eve's measurement from his post-selection. For squeezing parameter $r=1$ in Figure (\ref{jpd}), the maximum of joint probability results in $z_{1}=1.72$ and $x_{1}=0.28$, implying that if Bob measures a high value of his quadrature measurement $z_1$, Eve is likely to measure a small value of her quadrature measurement $x_1$. Hence, a high value of the post-selection threshold $z_c$ leads to high bit error rate in Eve's measurements, and therefore reducing Eve's information.

To estimate the performance of the suggested state for a given loss (eavesdropping or environment), we calculate the secure key gain per pulse ${G_{AB}}$. The maximum of secure key gain reveal the optimal values of post-selection threshold $z_c$ and the pulse intensity $n$ for the protocol.

The noisy channel is simulated by a beam-splitter with transmitivity $T$ and reflectivity $R$, where $T^2 +R^2 = 1$. The reflected part of the signal is parametrized by $({R^2}n)$ which represent the transmission loss, and the remaining part of the signal is received by Bob with $({T^2}n)$. The secure key gain is given by
\begin{equation}\label{key}
\displaystyle {G_{AB}}\left( {{z_c},n,r,{T^2}} \right) = \frac{1}{2}\Pi\left( {{z_c},{T^2}n,r} \right)\left( {{I_{AB}}\left( {{z_c},{T^2}n,r} \right) - \tau \left( {{(1-T^2)}n} \right)} \right)
\end{equation}
where
\begin{eqnarray}\label{key}
\displaystyle {I_{AB}}\left( {{z_c},n,r} \right) &=& \frac{1}{{2\Pi\left( { n ,{z_c}} \right)}}\int_{\left| {{z_1}} \right| > {z_c}} {{P_{{\psi _ + }}}\left( {{z_1},r,\sqrt n } \right){{\log }_2}\left( {{P_{{\psi _ + }}}\left( {{z_1},r,\sqrt n } \right)} \right)} \nonumber\\
&+& {{P_{{\psi _ - }}}\left( {{z_1},r,\sqrt n } \right){{\log }_2}\left( {{P_{{\psi _ - }}}\left( {{z_1},r,\sqrt n } \right)} \right)} \nonumber \\
&+& \left( {{P_{{\psi _ - }}}\left( {{z_1},r,\sqrt n } \right) + {P_{{\psi _ + }}}\left( {{z_1},r,\sqrt n } \right)} \right)\\ &\times&\left( {1 - {{\log }_2}\left( {{P_{{\psi _ - }}}\left( {{z_1},r,\sqrt n } \right) + {P_{{\psi _ + }}}\left( {{z_1},r,\sqrt n } \right)} \right)} \right){\rm{d}}{z_1}\nonumber
\end{eqnarray}
is the mutual information between Alice and Bob, which represents the survival information from the noisy channel. And $\tau \left( n \right) = 1 + {\log _2}\left( {{P_{coll}}\left( n \right)} \right)$ is the amount of reduction of the raw key where
\begin{equation}\label{key}
\displaystyle  {P_{coll}}\left( n \right) = \frac{1}{2}\mathop   \int_{ - \infty }^{ + \infty } \frac{{{P_{{\psi _ - }}}{{\left( {{z_1},r,\sqrt n } \right)}^2} + {P_{{\psi _ + }}}{{\left( {{z_1},r,\sqrt n } \right)}^2}}}{{{P_{{\psi _ - }}}\left( {{z_1},r,\sqrt n } \right) + {P_{{\psi _ + }}}\left( {{z_1},r,\sqrt n } \right)}}{\rm{d}}{{\rm{z}}_1}
\end{equation}
stands for the collision probability, which is significant in producing the secret key, because it indicates how much the raw key must be decreased to erase Eve's knowledge about it \cite{6,23}, and we define the accepted bit fraction as

\begin{equation}\label{key}
\displaystyle {r_{acc}}\left( {{z_c},r,n} \right) = \frac{1}{2}\left( {{P_0}\left( {{z_c},r,n} \right) + {P_1}\left( {{z_c},r,n} \right)} \right)
\end{equation}
where ${P_0}\left( {{z_c},r,n} \right)$ and $ {P_1}\left( {{z_c},r,n} \right)$ are the probabilities of measuring bit "0" and bit "1" respectively
\begin{equation}\label{key}
\begin{array}{l}
\displaystyle {P_0}\left( {{z_c},r,n} \right) = \int_{ - \infty }^{ - z_c}  {{P_{{\rho _1}}}\left( {{z_1},r,\sqrt n } \right){\rm{d}}{z_1}} ;\\
\displaystyle {P_1}\left( {{z_c},r,n} \right) = \int_{z_c}^{ + \infty } {{P_{{\rho _1}}}\left( {{z_1},r,\sqrt n } \right){\rm{d}}{z_1}} ;
\end{array}
\end{equation}
\begin{figure}[H]
	{\begin{minipage}[b]{0.48\linewidth}
			\centering
			\includegraphics[scale=0.3]{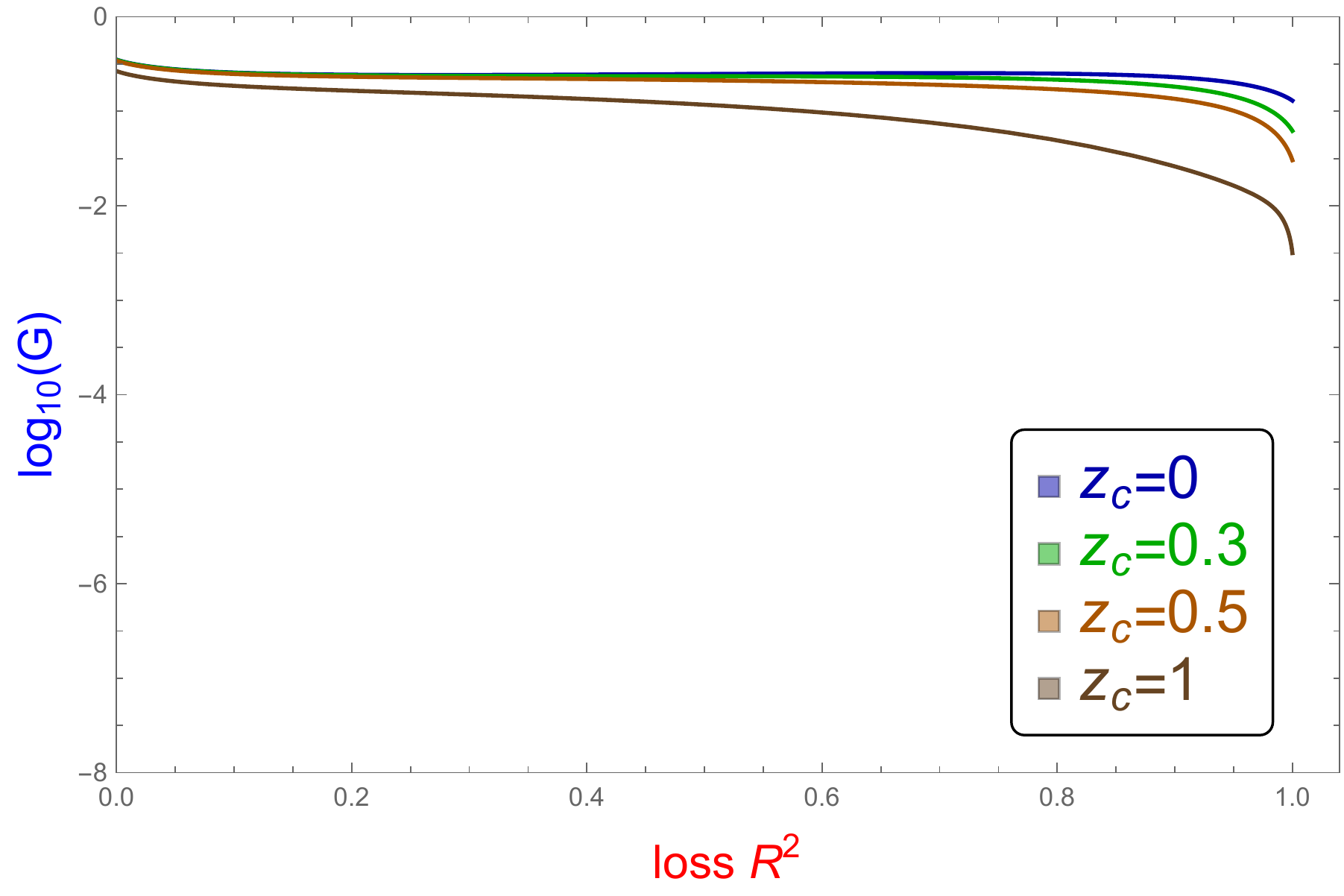}
			\caption{Secure key gain $G_{AB}$ as a function of the considered loss parameter
			$R^2$ for the pulse intensity $n = 1$, $r=0.2$ and the threshold values $z_c= 0, 0.3, 0.5, 1$.}
			\label{gain}
			\vfill     
		\end{minipage}\hfill
		\begin{minipage}[b]{0.48\linewidth}
			\centering
			\includegraphics[scale=0.3]{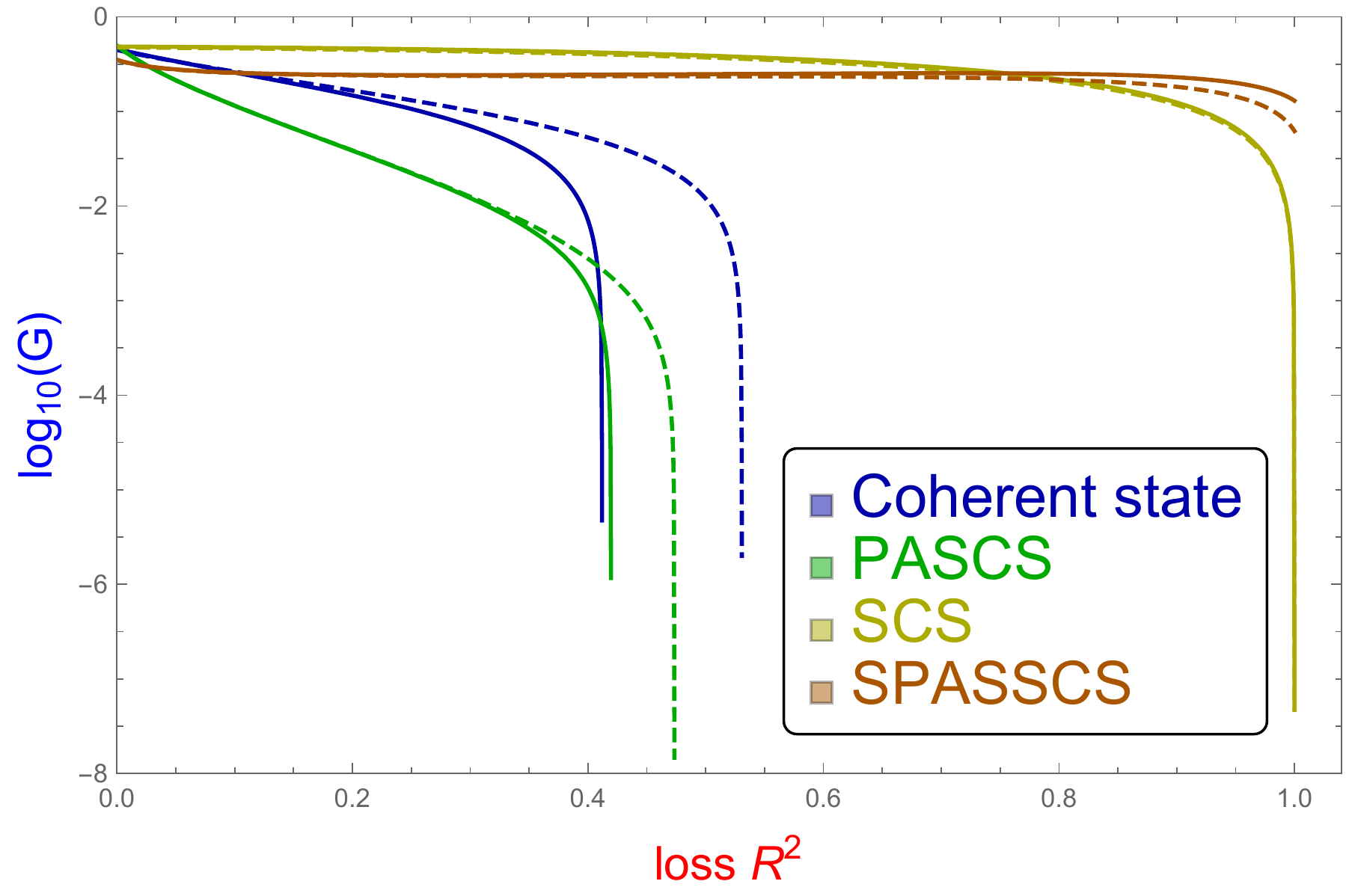}
			\caption{Secure key gain $G_{AB}$ of coherent state, PASCS, SCS and SPASSCS, for the threshold value $z_c =0$ solid lines and $z_c =0.3$ dashed lines, where the pulse intensity $n = 1$ and $r=0.2$.}
			\label{gainc}		\vfill     
	\end{minipage}}
\end{figure}

The effect of loss channel on the secure key gain $G_{AB}$ for different values of $z_c$ are depicted in Figure (\ref{gain}). The SPASSCS maintain the secure key gain $G_{AB}$ for different ratio of loss channel. It demonstrates a resistant state against loss channel than coherent state \cite{19}, SCS \cite{6} and PASCS \cite{12}, which vanish too quickly at certain loss channel ratios figure (\ref{gainc}) and necessitate raising the threshold value to hold the secure key gain much longer. This comparison boosts our suitable choice by confirming the robustness of our proposal state over others.

The maximum of $G_{AB}$ determines exactly the optimal coordinates of post-selection threshold value $z_c$ and pulse intensity $n$ for different loss ratio. For 90\% loss channel the maximum of secure key gain is ${G_{max}}=0.22$, which corresponds to optimal values $({z_c}=6.69.10^{-7}; n\simeq1)$, for 50\% loss channel we have ${G_{max}}=0.24$ corresponds to $({z_c}=5.86.10^{-6}; n\simeq1)$, and for 10\% loss channel, we have ${G_{max}}=0.31$ corresponds to $({z_c}=4.75.10^{-7}; n=0.14)$. As a summary of previous results, the SPASSCS does not require optimal coordinates of $n$ and $z_c$ values to achieve the security of the QKD protocol, and this is due to the convergence between optimal coordinates in different cases of the loss channel ratio. The latter is confirmed in figure (\ref{gain}), since whatever $z_c$ is valued, the secure key gain is withstood even in high loss channel ratio. Unlike other states, which require the optimal coordinates of $n$ and $z_c$ to ensure the security of the QKD protocol.

\begin{figure}[H]
	\centering
	\includegraphics[scale=0.3]{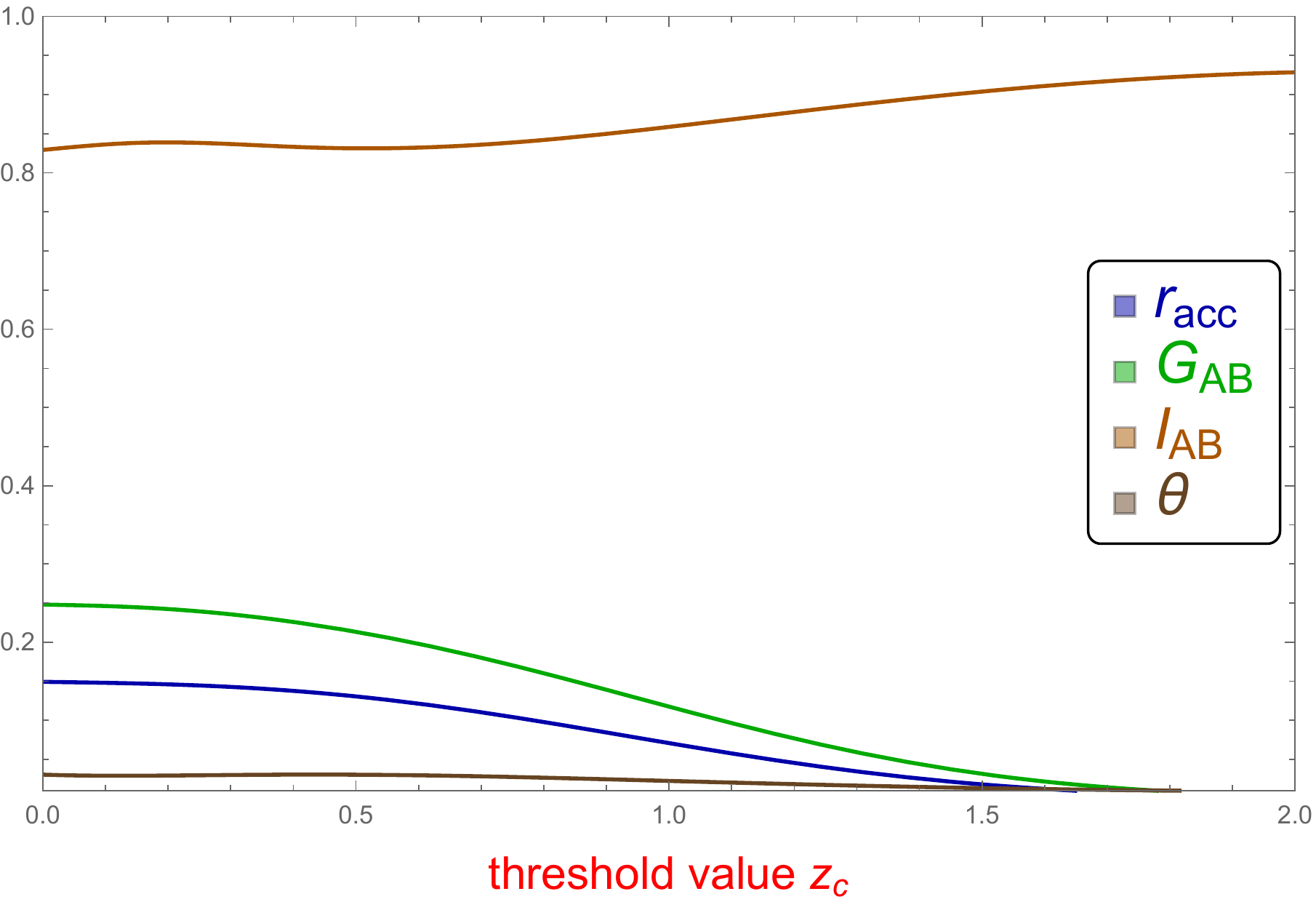}
	\caption{Significant quantities in function of threshold value $z_c$, with 50\% loss, $G_{AB}$, $I_{AB}$, $r_{acc}$ and $\Theta$ for the pulse intensity $n = 1$, $r=0.2$.}
	\label{report}
\end{figure}
The simultaneous behavior of some significants functions, fraction of accepted bits $r_{acc}$, bit error rate $\Theta$, mutual information $I_{AB}$, and secure key gain $G_{AB}$ as a function of threshold value $z_c$ in 50\% loss channel are shown in figure (\ref{report}). They are showing a similar behavior as reported in \cite{20} and \cite{6}. The interesting point here is the extremely low bit error rate. The results in figure (\ref{report}) can be interpreted as follow: when the threshold value $z_c$ increases, the post-selection will expands the error range (see figure (\ref{errange})), which makes BER $\Theta$ so small and increases the mutual information $I_{AB}$ between Alice and Bob. As results of this expansion, the post-selection will ignore most encoding pulses because of uncertainty interval's high width ]-$z_c$, $z_c$ [, which decreases the fraction of accepted bits $r_{acc}$ and in turn decreases the secure key gain $G_{AB}$. Thus, making the threshold value $z_c$ between 0 and 1, ensures the security of QKD protocol and maximizes the secure key gain $G_{AB}$ as mentioned in figure (\ref{gain}).

\section{Conclusion}
In this work we have investigated the BB84 continuous variables quantum key distribution protocol with post-selection and homodyne detection using single photon added then subtracted squeezed coherent state SPASSCS. We have shown the non Gaussian and non classical properties of the suggested state. We described the continuous variables BB84 protocol based on SPASSCS in ideal conditions without eavesdropping attacks. We examine Bob's bit error rate which shows a minimal error reducible by threshold parameter $z_c$.

Two kinds of eavesdropping strategies have been investigated. The first one is the intercept-resend attack, where Eve's intervention modifies the regular bases distributions, and increases the bit error rate on Bob's measurements. We show that the raise of the threshold value $z_c$ reduce Bob's bit error rate. The second eavesdropping strategies is the superior channel attack, where the maximum of joint probability shows a correlated beams and thus correlated measurements between Bob and Eve. This correlation gives Bob control on Eve's measurement by modifying his post-selection threshold $z_c$, and therefore minimizing Eve's knowledge about the secret key.

We have measured the secure key gain for a given loss. We prove that the SPASSCS maintain the information for high losses compared to coherent state \cite{19}, PASCS \cite{12} and SCS \cite{6}, then we describe the behavior of the state by displaying a simultaneous action of some significants quantities which resulting to the efficiency of the state against loss channel or attacks.

\end{document}